\documentclass[superscriptaddress,aps,journal=prl,preprint]{revtex4-2}
\usepackage{CJK}  

\usepackage{graphicx}
\usepackage{epstopdf}
\usepackage{bm,amsmath,amssymb} 
\usepackage{color, soul} 
\usepackage{dcolumn}   
\usepackage{multirow}
\usepackage{makecell}
\usepackage{threeparttable}
\usepackage{verbatim}
\newcommand{\note}[1]{{\color{black}{{#1}}}}

\newcommand{\eg}{{\em e.\,g.}}

\begin{document}
\begin{CJK*}{UTF8}{gbsn}

\title{A universal interatomic potential for perovskite oxides}

\author{Jing Wu ({\CJKfamily{gbsn}武静})}
\thanks{These authors contributed equally}
\affiliation{Key Laboratory for Quantum Materials of Zhejiang Province, Department of Physics, School of Science and Research Center for Industries of the Future, Westlake University, Hangzhou, Zhejiang 310030, China}
\author{Jiyuan Yang ({\CJKfamily{gbsn}杨季元})}
\thanks{These authors contributed equally}
\affiliation{Key Laboratory for Quantum Materials of Zhejiang Province, Department of Physics, School of Science and Research Center for Industries of the Future, Westlake University, Hangzhou, Zhejiang 310030, China}
\author{Yuan-Jinsheng Liu ({\CJKfamily{gbsn}刘袁今生})}
\thanks{These authors contributed equally}
\affiliation{Key Laboratory for Quantum Materials of Zhejiang Province, Department of Physics, School of Science and Research Center for Industries of the Future, Westlake University, Hangzhou, Zhejiang 310030, China}
\author{Duo Zhang ({\CJKfamily{gbsn}张铎})}
\affiliation{DP Technology, Beijing 100080, China}
\affiliation{AI for Science Institute, Beijing 100080, China}
\author{Yudi Yang ({\CJKfamily{gbsn}杨雨迪})}
\affiliation{Key Laboratory for Quantum Materials of Zhejiang Province, Department of Physics, School of Science and Research Center for Industries of the Future, Westlake University, Hangzhou, Zhejiang 310030, China}
\author{Yuzhi Zhang ({\CJKfamily{gbsn}张与之})}
\affiliation{DP Technology, Beijing 100080, China}
\affiliation{AI for Science Institute, Beijing 100080, China}
\author{Linfeng Zhang ({\CJKfamily{gbsn}张林峰})}
\affiliation{DP Technology, Beijing 100080, China}
\affiliation{AI for Science Institute, Beijing 100080, China}
\author{Shi Liu ({\CJKfamily{gbsn}刘仕})}
\email{liushi@westlake.edu.cn}
\affiliation{Key Laboratory for Quantum Materials of Zhejiang Province, Department of Physics, School of Science and Research Center for Industries of the Future, Westlake University, Hangzhou, Zhejiang 310030, China}
\affiliation{Institute of Natural Sciences, Westlake Institute for Advanced Study, Hangzhou, Zhejiang 310024, China}


\begin{abstract}{ 
With their celebrated structural and chemical flexibility, perovskite oxides have served as a highly adaptable material platform for exploring emergent phenomena arising from the interplay between different degrees of freedom. Molecular dynamics (MD) simulations leveraging classical force fields, commonly depicted as parameterized analytical functions, have made significant contributions in elucidating the atomistic dynamics and structural properties of crystalline solids including perovskite oxides. 
However, the force fields currently available for solids are rather specific and offer limited transferability, making it time-consuming to use MD to study new materials systems since a new force field must be parameterized and tested first. 
The lack of a generalized force field applicable to a broad spectrum of solid materials hinders the facile deployment of MD in computer-aided materials discovery (CAMD). Here, by utilizing a deep-neural network with a self-attention scheme,
we have developed a unified force field (UniPero) that enables MD simulations of perovskite oxides involving 14 metal elements and conceivably their solid solutions with arbitrary compositions.  Notably, isobaric-isothermal ensemble MD simulations with this model potential accurately predict the experimental temperature-driven phase transition sequences for several markedly different ferroelectric oxides, including a 6-element ternary solid solution, Pb(In$_{1/2}$Nb$_{1/2}$)O$_3$--Pb(Mg$_{1/3}$Nb$_{2/3}$)O$_3$--PbTiO$_3$. We believe the universal interatomic potential along with the training database, proposed regression tests, and the auto-testing workflow, all released publicly, will pave the way for a systematic improvement and extension of a unified force field for solids, potentially heralding a new era in CAMD. 
}
\end{abstract}

\maketitle
\end{CJK*}

\newpage
Classical force fields, including CHARMM~\cite{Brooks83p187}, AMBER~\cite{Case05p1668}, and GROMOS~\cite{Scott99p3596}, which are capable of accurately describing the interatomic interactions within key biological molecules such as amino acids and nucleic acids, have played a pivotal role in expanding the use of molecular dynamics (MD) simulations in computer-aided drug design (CADD). These force fields enabled large-scale simulations of atomic and molecular movements in biological systems with remarkable efficiency and precision, greatly facilitating detailed studies of drug-receptor interactions~\cite{Pan13p667}, protein conformational changes~\cite{Karplus02p646,Shaw10p341}, binding affinities~\cite{Aaqvist02p358}, and the search for lead compounds~\cite{Sakkiah12p3267}; they are now indispensable tools in the pursuit of new drugs and therapeutics~\cite{Durrant11p1,Lemkul12p845}. In contrast, although MD simulations have been successful in elucidating the properties of various types of solids by offering atomic-level insights, their integration into computer-aided materials discovery (CAMD) has not yet achieved the same level of industry-wide adoption as in CADD. This disparity is primarily due to the lack of a generalized force field applicable to a wide range of solid materials.

Many force fields for crystalline solids are tailored for a specific material or a narrow group of materials, covering a limited chemical space~\cite{Ohira94p565,Sepliarsky05p107, Shin05p054104, Sepliarsky11p435902,Wexler19p174109,Sheng11p134118,Noordhoek13p274,Richard22p054066}.
While the reactive force-field (ReaxFF) interatomic potential allows for element-specific parameterization and is commonly viewed as possessing strong transferability~\cite{vanDuin01p9396}, its developers still caution against arbitrarily combining parameter sets in anticipation of consistent predictive power~\cite{Senftle16p15011}. Moreover, the usage of a highly sophisticated energy  function containing a large number of empirical parameters (as seen in ReaxFF) makes it challenging to parameterize force fields for new materials. 
In this work, taking $AB$O$_3$-type perovskite oxides as an example, we demonstrate that a deep neural network-based  model, namely deep potential (DP) ~\cite{Zhang18p143001} with a novel attention mechanism, referred to as DPA-1~\cite{zhang2022dpa1}, exhibits the required representability and transferability for the development of a universal interatomic potential. By adopting the strategy of ``modular development of deep potential" (ModDP)~\cite{Wu23p144102} and relying solely on {\em ab initio} training data, we have achieved a 
unified force field that enables MD simulations for diverse perovskite oxides comprising 14 different metal elements and is conceivably adaptable for their solid solutions of any composition. For instance, we report the first large-scale MD simulations of a 6-element ternary solid solution, Pb(In$_{1/2}$Nb$_{1/2}$)O$_3$--Pb(Mg$_{1/3}$Nb$_{2/3}$)O$_3$--PbTiO$_3$ (PIN-PMN-PT), which is a relaxor ferroelectric possessing superior piezoelectric properties~\cite{Zhang08p064106, Liu09p074112}. 

The $AB$O$_3$-type perovskite oxide is chosen as the model system because of its role as an exceptionally versatile material platform. 
The adaptability in chemical, structural, and compositional variations grants a unique capability to fine-tune interactions such as electron-phonon coupling, crystal field, and spin-orbit interaction, the competitions among which often dictate the functional properties of perovskite oxides~\cite{Yi17p443004, Ramesh19p257}. Given these intricacies, it becomes challenging, if not outright impractical, to rely on a force field with a fixed analytical structure and a limited set of parameters to create a generalized force field for perovskite oxides spanning a broad chemical and compositional range. To address this issue, we adopt the DP model and its extension augmented with a self-attention scheme.

The DP model, based on a deep neural network with the number of learnable parameters on the order of 10$^{6}$, offers a robust mathematical structure to represent highly nonlinear and complex interatomic interactions while bypassing the need to handcraft descriptors that represent local atomic environments~\cite{Zhang18p143001}. Specifically, the DP model features a symmetry-preserving embedding network that maps an atom's local environment to inputs for a fitting neural network which then outputs the atomic energy; the sum of atomic energies yields the total energy.
\note{This approach is a manifestation of the embedded atom concept, capturing the many-body character of interatomic interactions. Therefore, the DP model trained using DFT total energies implicitly accounts for various types of interactions including the long-range interactions.}
More recently, the DPA-1 model~\cite{zhang2022dpa1} has introduced an element-type embedding net and integrated a self-attention mechanism. This mechanism excels in modeling the importance of neighboring atoms and reweighting the interactions among them based on both distance and angular information, thus allowing for the mixing and communication within the latent space of the embedding parameters of elements and structures.
As a result, the DPA-1 model achieves satisfactory transferability to unseen systems with varied elemental compositions~\cite{zhang2022dpa1}, fulfilling the essential criteria for constructing a universal force field applicable to solids comprising multiple elements.

To develop a DPA-1 model capable of representing diverse element types and compositions that the perovskite structure can support, the construction of a comprehensive training database becomes pivotal.
For simple materials systems comprising two or three elements, the DP-GEN scheme~\cite{Zhang19p023804, Zhang20p107206} for data generation is streamlined and largely autonomous, requiring little human involvement. This concurrent learning procedure iteratively explores the configuration space (by running MD simulations) using one of the four models trained with existing data. It then labels only those MD-sampled configurations that exhibit high uncertainty levels, determined by the
maximum standard deviation of predictions from the four models. The energies and forces of these newly labeled configurations, computed using first-principles density functional theory (DFT), are subsequently incorporated into the training database for the next learning cycle. The iteration stops when all configurations sampled from MD simulations meet a certain standard of accuracy across all four models. More details regarding DP-GEN can be found in previous studies~\cite{Zhang20p107206, Wu21p024108, Wu21p174017}. However, the robustness of the DP-GEN scheme is limited when constructing the training database for complex oxides with diverse element types and compositions.

\begin{figure}[]
\centering
\includegraphics[width=1.0\textwidth]{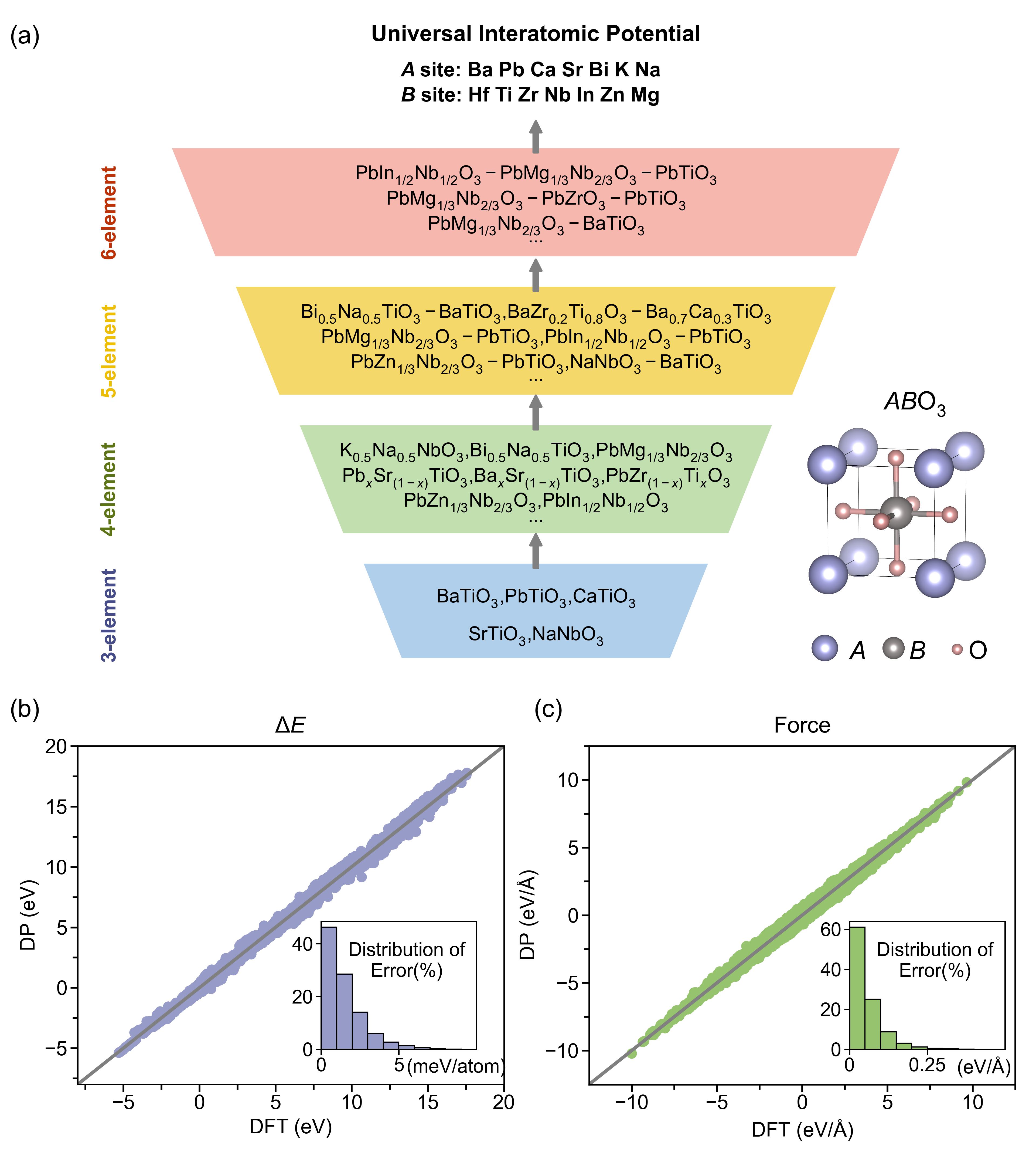}
\caption{(a) Workflow for the development of a universal force field of perovskite oxides following the ModDP protocol. Comparison of (b) relative energies ($\Delta E$) and (c) atomic forces predicted using the DPA-1 model with reference DFT results for $\approx$20,000 configurations in the final training database. The inset shows the distribution of the absolute error.
}
\label{Fitting}
\end{figure}

In previous work, we demonstrated that the ModDP protocol facilitates a systematic development and improvement of DP models for complex oxide solid solutions~\cite{Wu23p144102}.
At its core, ModDP embodies the concepts of ``data reusing" and ``divide and conquer": the converged training database associated with an end-member material is treated as an independent module, and is then reused to train the DP model of solid solutions derived from these end-members. Following the same spirit, we devise a procedure that progressively introduces perovskites with increasing complexity (number of element types) to enhance the capability of DPA-1 level by level. The workflow of the force field development procedure that leads to a universal interatomic potential is depicted in Fig.~\ref{Fitting}a. The initial database contains $\approx$1000 configurations of 26 different types of perovskites, involving $\approx$200 compositions and 14 metal elements. It is noted that some data for PbTiO$_3$ and SrTiO$_3$ were taken from a published database~\cite{Wu23p144102}, and we did {\em not} intentionally design the initial database, aiming to minimize human intervention. We first use the standard DP-GEN scheme to converge a DPA-1 model suited for 3-element (including oxygen) perovskite oxides such as PbTiO$_3$, BaTiO$_3$, CaTiO$_3$, SrTiO$_3$, and NaNbO$_3$. Subsequently, the converged training database serves as the starting point for DP-GEN to improve DPA-1 for 4-element perovskite systems like PbZr$_{1-x}$Ti$_{x}$O$_3$ and Pb(Mg$_{1/3}$Nb$_{2/3})$O$_3$. Ultimately, we achieve a converged DPA-1 model for 6-element perovskite systems, including the ternary solid solution of PIN-PMN-PT. As detailed below, the final DPA-1 model serves as a universal interatomic potential (named as UniPero), capable of modeling a wide variety of perovskites with MD simulations.

All DFT calculations are performed with Atomic-orbital Based Ab-initio Computation at UStc (\texttt{ABACUS}) package~\cite{Chen10p445501,Li16p503} using numerical atomic orbitals (NAOs) and SG15 optimized norm-conserving Vanderbilt (ONCV) pseudopotentials~\cite{Schlipf15p36}. The Perdew-Burke-Ernzerhof revised for solids (PBEsol)~\cite{Perdew08p136406} within the generalized gradient approximation is chosen as the exchange-correlation functional, and the double-$\zeta$ plus polarization functions (DZP) with a plane-wave cutoff energy of 100 Ry is employed as the NAO basis set. The energies and forces are computed with a 0.1 Bohr$^{-1}$ $k$-point spacing and an energy convergence threshold of 1.0$\times 10^{-7}$ Ry.  All isobaric-isothermal ($NPT$) ensemble MD simulations are performed using \texttt{LAMMPS}~\cite{Plimpton95p1}, with a time step of 2~fs and the temperature controlled via the Nos\'e-Hoover thermostat and the pressure controlled by the Parrinello-Rahman barostat. \note{When simulating temperature-driven phase transitions, the perovskite systems are modeled with a 10$\times$10$\times$10 supercell (assuming a 5-atom unit cell) consisting of 5,000 atoms, except for PIN-PMN-PT which is modeled with a 20$\times$20$\times$20 supercell consisting of 40,000 atoms.} The smooth edition of the DP model (with a cutoff radius of 6~\AA~and the smoothing starting at 0.5~\AA) with an attention scheme, DPA-1, is adopted. The \texttt{DEEPMD-KIT} package~\cite{Wang18p178} is used in the training cycle of DP-GEN. 
The basic networks of DPA-1 comprise the embedding and fitting networks, both of which utilize ResNet architectures with default settings. Additionally, DPA-1 also incorporates a one-hot type embedding network with a size of 8, as well as a two-layer attention mechanism possessing 128 hidden dimensions during the scaled-dot product attention computation.
To enhance reproducibility, we have made our final training database and hyperparameters accessible via a public repository~\cite{BohriumNotebook}.

The final training database contains 19288 configurations. We note that DPA-1 aims to reproduce DFT total energies, which inherently vary significantly between different materials. For configurations in the database, the DFT total energies span a range of an astonishing $\approx$400~eV/atom (including an element-specific constant energy shift). Impressively, the DPA-1 model has a mean absolute error (MAE) of 1.75 meV/atom for the total energy, highlighting its remarkable representability. To better illustrate the fitting performance, we compare the energy differences ($\Delta E$) predicted by DPA-1 with DFT values. For a specific composition, a configuration is (arbitrarily) chosen as the reference; the $\Delta E$ values are then computed relative to the total energy of this chosen reference configuration. As shown in Fig.~1b, DPA-1 well reproduces DFT values of $\Delta E$. The predictions from DPA-1 for atomic force are in satisfactory agreement with DFT results as well, achieving a MAE of 0.054 eV/\AA~(Fig.~1c).

Employing a concurrent learning approach like DP-GEN allows the training database to be updated efficiently and iteratively. As a result, we do not have the typical validation or testing datasets commonly associated with conventional supervised learning. We propose two sets of tasks to evaluate DPA-1's applicability for perovskite oxides. For the first set of testing jobs (denoted as Task I), AIMD simulations are performed at three specific temperatures (300, 450, 900~K) for a few picoseconds. DPA-1 is then employed to evaluate the energies for configurations along these AIMD trajectories. The objective is to ascertain DPA-1's accuracy in predicting energies for configurations sampled by AIMD at finite temperatures.
As illustrated in Fig.~2a which uses PIN-PMN-PT as an example, the energy evolution trajectory obtained with DPA-1 closely follows that from AIMD at three different temperatures, confirmed by Pearson correlation coefficients of 0.93, 0.95, and 0.99, respectively. 
The second set of evaluations (Task II) is to generate classical DPMD trajectories using DPA-1 at varying temperatures, the energies of configurations along which are subsequently determined with DFT. This task is designed to validate whether classical MD trajectories from DPA-1 properly sample the relevant configuration spaces at finite temperatures.
The applicability of DPA-1 is gauged by the MAEs of energies and atomic forces for all the configurations generated from these two sets of tasks. We have selected 10 representative perovskite oxides for testing: BaTiO$_3$, PbTiO$_3$, SrTiO$_3$, Ba$_{0.5}$Sr$_{0.5}$TiO$_3$, K$_{0.5}$Na$_{0.5}$NbO$_3$, NaNbO$_3$--BaTiO$_3$, PbZr$_{0.5}$Ti$_{0.5}$O$_3$, BaZr$_{0.2}$Ti$_{0.8}$O$_3$--Ba$_{0.3}$Ca$_{0.7}$TiO$_3$, PMN--PT, and PIN--PMN--PT. It is evident from Fig.~2c (for Task I) and Fig.~2d (for Task II) that the DPA-1 model well reproduces the energy evolution from AIMD and performs appropriate sampling at finite temperatures for markedly different perovskite systems.  The MAE for energy is lower than 1.45 meV/atom, and that for atomic force is below 0.052 eV/\AA.

With DPA-1's capability to faithfully capture the complex and highly nonlinear potential energy surfaces involving diverse element types, we expect developing a DPA-1 model with further enhanced transferability and accuracy can be achieved by introducing more data from new materials systems into the current training database. Therefore, the proposed tasks, which are readily automatable, serve as suitable {\em regression tests} (a common practice in software development) for the continuous improvement of the DPA-1 model. We have implemented an auto-testing workflow based on an online platform (Bohrium Notebook)~\cite{BohriumNotebook} that allows real-time access to our DPA-1 model, cloud-based computing resources, and data analysis scripts. This initiative aims to streamline force field development and foster collaborative contributions from the community.

A definitive test of the applicability of the developed DPA-1-based universal interatomic potential is to examine temperature-driven phase transitions in ferroelectric perovskites. As summarized Fig.~\ref{PhaseTrans}, a single DPA-1 model, without any tuning, successfully reproduces the experimental sequences of phase transitions for lead-based ferroelectrics (PbTiO$_3$, Pb$_{0.5}$Sr$_{0.5}$TiO$_3$, and PbZr$_{0.5}$Ti$_{0.5}$O$_3$), lead-free ferroelectrics (BaTiO$_3$, KNbO$_3$, and K$_{0.5}$Na$_{0.5}$NbO$_3$), quantum paraelectric SrTiO$_3$, and tertiary solid solutions (0.29PIN--0.45PMN--0.26PT and 0.36PIN--0.36PMN--0.28PT). For example, the temperature-dependent lattice constants obtained from $NPT$ MD simulations reveal a tetragonal ($T$, space group $P4mm$) to cubic ($C$, space group $Pm\bar{3}m$) transition at $\approx$550~K in PbTiO$_3$ (Fig.~\ref{PhaseTrans}a), whereas the transition temperature ($T_c$) drops to $\approx$280~K in the $A$-site doped solid solution, Pb$_{0.5}$Sr$_{0.5}$TiO$_3$ (Fig.~\ref{PhaseTrans}d). In comparison, MD simulations using DPA-1 predict that the $B$-site doped solid solution, PbZr$_{0.5}$Ti$_{0.5}$O$_3$, 
adopts a rhombohedral phase ($R$, space group $R3m$) at low temperatures, and it undergoes rhombohedral-tetragonal-cubic ($R$-$T$-$C$) phase transitions as the temperature increases (Fig.~\ref{PhaseTrans}g), consistent with experimental observations~\cite{Jaffe54p809,JAFFE71}. SrTiO$_3$ is a well-known quantum paraelectric that exhibits a temperature-driven tetragonal-cubic transition ($I4/mcm\rightarrow Pm\bar{3}m$) characterized by a decrease in the TiO$_6$ octahedral tilt angle. This phase transition, corresponding to a tiny energy barrier of $\approx$1 meV/atom, is reproduced by DPA-1 (Fig.~3b), and the theoretical value of $T_c$ is 150~K, which is comparable to the experimental value of 105~K~\cite{Riste71p1455}. The same model also correctly predicts the phase transition sequence, rhombohedral-orthorhombic-tetragonal-cubic ($R$-$O$-$T$-$C$), in BaTiO$_3$, KNbO$_3$, and K$_{0.5}$Na$_{0.5}$NbO$_3$ (see Fig.~\ref{PhaseTrans}c, e, and h). Moreover, even for the highly complex 6-element tertiary solid solutions such as PIN-PMN-PT, DPA-1 not only reproduces the $T$--$C$ phase transition observed in experiments, but also predicts the correct trend in the change of $T_c$ due to the compositional variation. As shown in Fig.~\ref{PhaseTrans}c and d, the $T_c$ rises by about 50~K from 0.29PIN--0.45PMN--0.26PT to 0.36PIN--0.36PMN--0.28PT, similar to the $T_c$ change of $\approx$60~K found in experiments due to the increased percentage of PIN from 29\% to 36\%~\cite{Li18p345,Wang12p433}. It is observed that theoretical $T_C$ values for ferroelectric transitions are typically lower than experimental values by 100-200~K. Such discrepancies appear to be a common feature for force fields developed by fitting to results computed with the PBEsol density functional~\cite{Liu13p104102,Wu23p144102,Qi16p134308}. Given the high fidelity of DPA-1 as demonstrated by its excellent fitting to DFT results, the underestimation of $T_C$ is more likely a reflection of PBEsol's limitations in forecasting finite-temperature properties rather than an issue with DPA-1's precision. 

The universal interatomic potential developed here can serve as a readily deployable tool to reveal the local structures and lattice dynamics of perovskite oxides with atomic spatial resolution. Using three 50-ps equilibrium DPMD trajectories for 0.29PIN--0.45PMN--0.26PT, each with different $B$-site cation arrangements and modeled with a 40,000-atom supercell, we analyze the distributions of local displacements of cations with respect to the center of their surrounding oxygen cages. \note{For describing inversion symmetry breaking in ferroelectric materials, polarization is often the natural order parameter. 
Given a configuration from MD simulations, 
the polarization for simple perovskite systems such as PbTiO$_3$ can be estimated using the following formula,
\[\mathbf{P}^m(t)=\frac{1}{V_{\rm uc}}\left[\frac{1}{8} \mathbf{Z}_{\mathrm{Pb}}^* \sum_{i=1}^8 \mathbf{r}_{\mathrm{Pb}, i}^m(t)+\mathbf{Z}_{\mathrm{Ti}}^* \mathbf{r}_{\mathrm{Ti}}^m(t)+\frac{1}{2} \mathbf{Z}_{\mathrm{O}}^* \sum_{i=1}^6 \mathbf{r}_{\mathrm{O}, i}^m(t)\right]\],
where $\mathbf{P}^m(t)$ is the polarization of unit cell $m$ at time $t$, $V_{\rm uc}$ is the volume of the unit cell, $\mathbf{Z}_{\mathrm{Pb}}^*, \mathbf{Z}_{\mathrm{Ti}}^*$, and $\mathbf{Z}_{\mathrm{O}}^*$ are the Born effective charges of Pb, Ti and O atoms, $\mathbf{r}_{\mathrm{Pb}, i}^m(t), \mathbf{r}_{\mathrm{Ti}, i}^m(t)$, and $\mathbf{r}_{\mathrm{O}, i}^m(t)$ are the instantaneous atomic positions. The polarization is thus the dipole moment per unit volume.} However, PIN-PMN-PT is considered a charge-frustrated system where the $B$ sites host cations of different charges: Mg$^{2+}$, In$^{3+}$, Ti$^{4+}$, and Nb$^{5+}$. \note{It becomes problematic to define a local electric dipole for a unit cell that is not locally charge neutral (\eg, a local PbInO$_3$ unit cell in PIN-PMN-PT solid solutions). Therefore, we use local atomic displacements of cations to measure the degree of inversion symmetry breaking at the unit cell level.}

In Fig.~4a-c, the probability density distributions of cation displacements ($d_x$, $d_y$, $d_z$) along the Cartesian axes are shown as a function of temperature. At a low temperature of 200~K, the distributions of $d_x$ and $d_y$ are mostly Gaussian-like and centered around zero for all cations. In contrast, the peak position for $d_z$ of Pb deviates the most away from zero, followed by Ti, Nb, In, and Mg in sequence.  This deviation in the $d_z$ distribution, compared to $d_x$ and $d_y$, aligns with the global tetragonal symmetry indicated by the lattice constants. Importantly, these distributions highlight a distinct difference in the local structures between relaxor ferroelectric PIN--PMN--PT and prototypical ferroelectrics like PbTiO$_3$. In PIN--PMN--PT, the distributions of cation displacements are more dispersive and isotropic, spanning both positive and negative values along all three Cartesian directions. Conversely, both Pb and Ti cations in ferroelectric PbTiO$_3$ displace along the same polar direction (\eg, [001])~\cite{Liu13p104102}. At an elevated temperature of 300~K, the peak positions of the $d_z$ distributions for all cations shift toward values lower than those at 200~K. When temperature rises above $T_C$, distributions become Gaussian, all centering at zero.

Snapshots of instantaneous cation displacements in configurations sampled from MD are displayed in \note{Fig.~4d-e}. We find that the magnitude of local Pb displacements exhibits less temperature sensitivity compared to the  $B$-site cations. Even at 400~K which is above the theoretical $T_C$, many Pb atoms remain locally displaced; their displacements grow more chaotic, manifesting diminished long-range order. The displacements of $B$-site cations, however, display more vulnerability: some are already nonpolar at 200~K, and an increasing percentage becomes nonpolar as temperature increases. Notably, we do not observe any polar nanoregions (PNRs) embedded in a nonpolar matrix, as the majority of Pb atoms are displaced, in agreement with previous MD simulations of PMN--PT~\cite{Takenaka17p391}.
The markedly different behaviors between Pb and $B$-site cations could be important for understanding the piezoelectric response of lead-based relaxor ferroelectrics. These findings underscore the quantitative nature of the DPA-1 model.

\note{We briefly comment on the main limitations and challenges of machine-learning force fields like DPA-1, which center around interpretability and computational efficiency. While the large number of learnable parameters of DPA-1 facilitates the representation of complex interactions, it simultaneously hinders our understanding of these very interactions.  
In essence, each of these parameters adjusts itself during the learning process to minimize prediction errors, but the collective behavior of such a high number of parameters can be enigmatic. This contrasts with traditional force fields which, through their explicit functional forms, often allow for clearer physical interpretation. Although the DPA-1 model demonstrates a clear speed advantage over first-principles methods like DFT, it still lags behind conventional force fields. For instance, the bond-valence model~\cite{Liu13p104102} typically runs two orders of magnitude faster than the DPA-1 model. 
Solutions to the aforementioned challenges exist.  Principal component analysis can be performed to analyze parameters of the DPA-1 model to explore the information within the latent space. For example, in a recent work of DPA-1~\cite{zhang2022dpa1}, the visualizations of the element-related parameters learned by the model coincide with the periodic table, partially demonstrating the model's interpretability. As for computational efficiency, model compression~\cite{Lu2021dpcompress} is set to alleviate this problem by using a small number of polynomial coefficients to fit the learned parameters in the model. This approach can significantly reduce the computational cost while maintaining accuracy.}

The current state of DPA-1 for perovskite oxides can be viewed as a pre-trained model, which is already sufficiently accurate for many perovskite oxides as discussed above. Drawing parallels from the development history of ChatGPT~\cite{Ouyang22p27730} in natural language processing, such a pre-trained model can be fine-tuned for specific downstream tasks with minimal extra effort using few-shot or zero-shot learning techniques~\cite{Wang20p1}. As the community continues to amass high-quality electronic structure data, it is conceivable that this pre-trained model will progressively expand its coverage, encompassing an ever broader range of chemical and structural spaces.


In summary, using the $AB$O$_3$-type perovskite oxide as a model system, our results demonstrate that a deep neural network, augmented with an attention scheme, can deliver the requisite representability and transferability to establish a universal interatomic potential for solid materials spanning diverse element types and compositions, at least for materials systems with a fixed crystal structure. Even with the complex nature of target systems involving 15 elements, the final training database remains impressively concise, containing no more than 20,000 configurations; the associated computational cost is manageable, requiring fewer than 200,000 CPU hours. The DPA-1-based force field, UniPero, developed in this work already exhibits satisfactory accuracy, proving suitable for large-scale MD simulations of distinct ferroelectric perovskites and reproducing the correct phase transition sequences driven by temperature. By publicly releasing the training database and the automated testing workflow, we anticipate our initiative will
facilitate a systematic improvement and extension of a generalized force field with an even larger scope \note{(\eg, free surfaces and defects)}. This, in turn, hopefully could foster the routine and efficient use of molecular dynamics for simulating solid materials, ultimately providing valuable insights for computer-aided materials discovery.

\begin{acknowledgments}
This work is supported by Natural Science Foundation of Zhejiang Province (2022XHSJJ006), National Natural Science Foundation of China (12074319), and Westlake Education Foundation. The computational resource is provided by Westlake HPC Center and Bohrium platform.
\end{acknowledgments}

\bibliography{SL}

\clearpage
\newpage

\clearpage
\newpage
\begin{figure}[]
\centering
\includegraphics[width=1.0\textwidth]{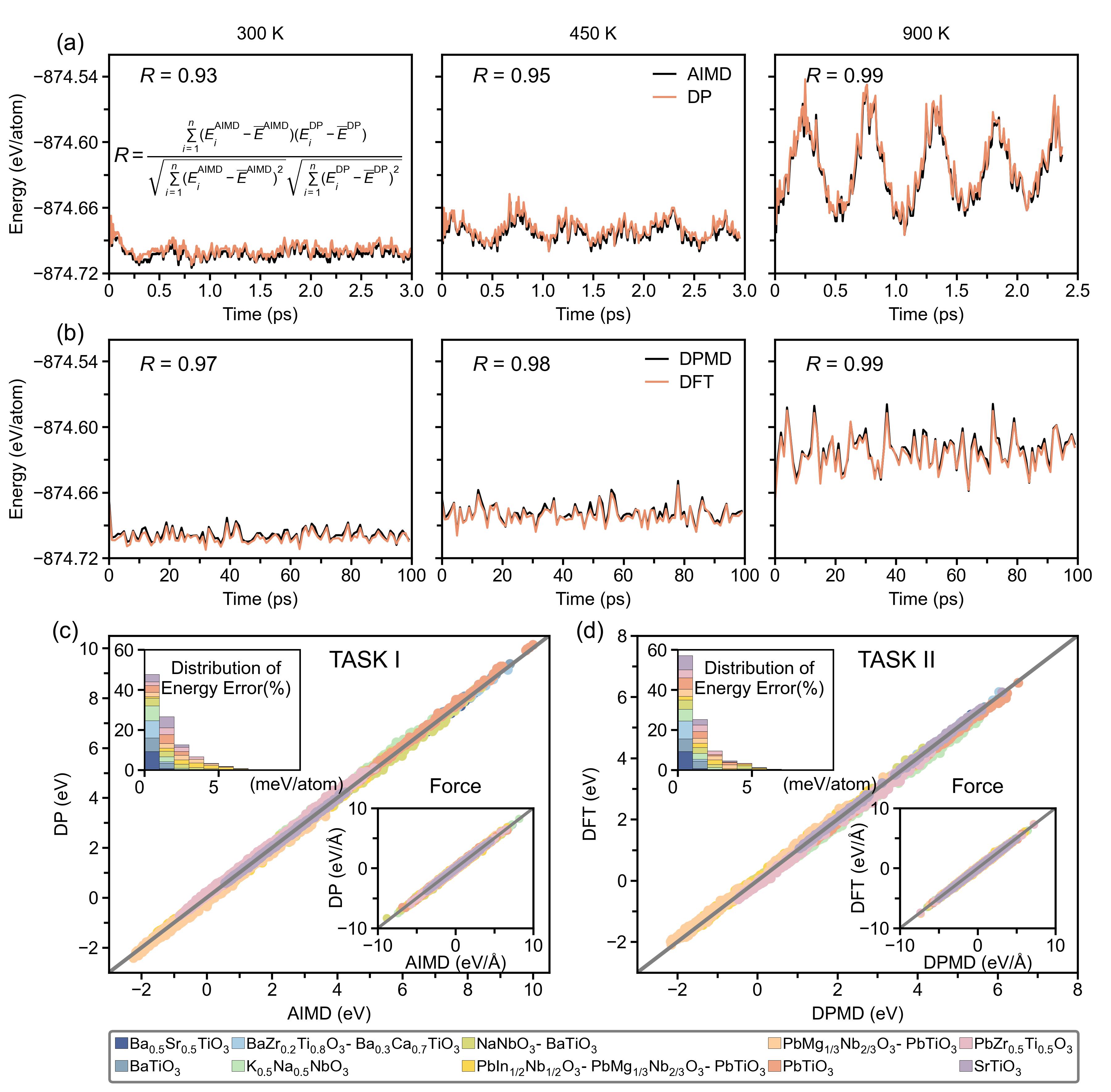}
\caption{Proposed tasks to evaluate the applicability of DPA-1 using PIN--PMN--PT as an example. (a) Task I. The energies of configurations sampled by AIMD at three different temperatures are checked by DPA-1. (b) Task II. The energies of configurations sampled by classical MD with DPA-1 are checked by DFT.  Comparison of (c) relative energies and (d) atomic forces for configurations of 10 representative perovskite oxides generated from Task I and Task II.
}
\label{Test}
\end{figure}

\clearpage
\newpage
\begin{figure}[]
\centering
\includegraphics[width=1.0\textwidth]{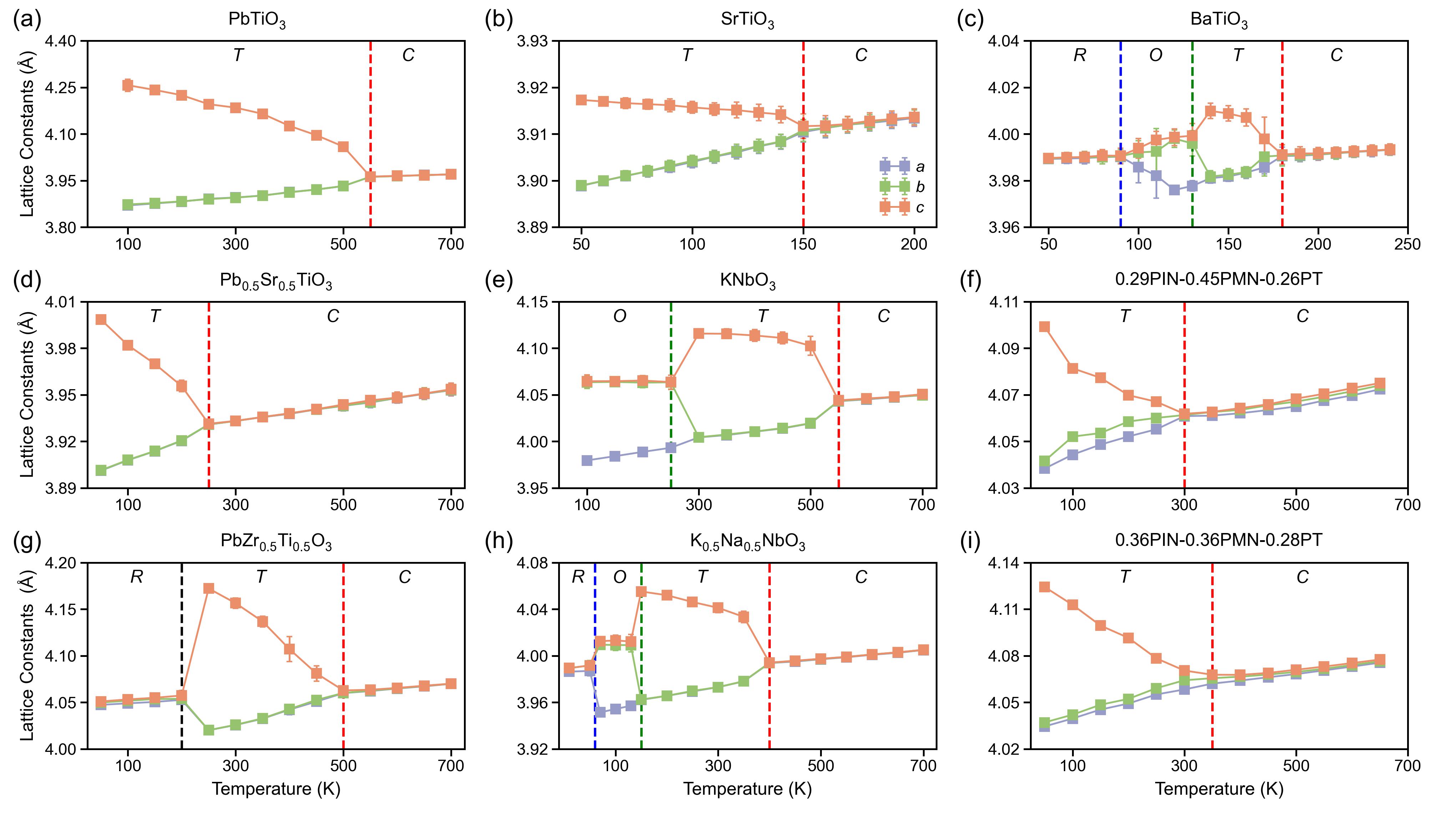}
\caption{Temperature-dependent lattice constants simulated with $NPT$ DPMD using the universal interatomic potential for (a) PbTiO$_3$, (b) SrTiO$_3$, (c) BaTiO$_3$, (d) Pb$_{0.5}$Sr$_{0.5}$TiO$_3$, (e) KNbO$_3$, (f) 0.29PIN--0.45PMN--0.26PT, (g) PbZr$_{0.5}$Ti$_{0.5}$O$_3$,  (h) K$_{0.5}$Na$_{0.5}$NbO$_3$, (i) 0.36PIN--0.36PMN--0.28PT.
}
\label{PhaseTrans}
\end{figure}

\clearpage
\newpage
\begin{figure}[]
\centering
\includegraphics[width=1.0\textwidth]{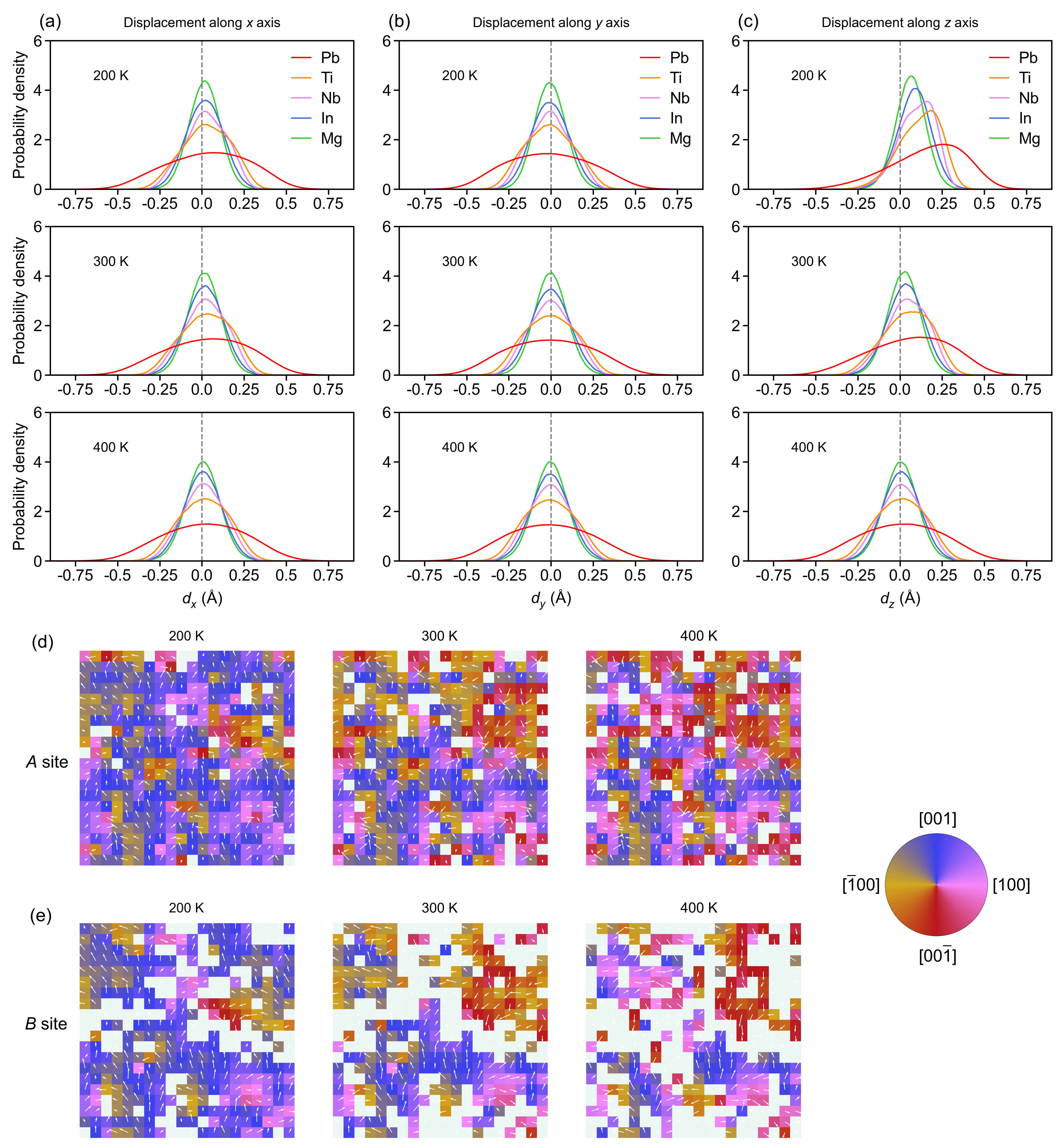}
\caption{Probability density distributions of cation displacements along the (a) $x$-axis, (b) $y$-axis, (c) $z$-axis as temperature increases in 0.29PIN--0.45PMN--0.26PT modeled with a 40,000-atom supercell. Snapshots of (d) $A$-site and (e) $B$-site cation displacements at different temperatures obtained
from MD simulations. Each white arrow indicates the local atomic displacement within a unit cell, with its background color representing the direction. Note that if the magnitude of the local displacement is less than 0.1~\AA, the background is colored white.
}
\label{PINPMNPT}
\end{figure}

\end{document}